\documentstyle[12pt,epsf,draft]{article}
\newcommand{\be}{\begin{equation}}
\newcommand{\ee}{\end{equation}}
\newcommand{\bea}{\begin{eqnarray}}
\newcommand{\eea}{\end{eqnarray}}
\newcommand{\ba}{\begin{array}}
\newcommand{\ea}{\end{array}}

\def\bbox{{\,\lower0.9pt\vbox{\hrule \hbox{\vrule height 0.2 cm
\hskip 0.2 cm \vrule height 0.2 cm}\hrule}\,}}
\newcommand{\dsl}{\pa \kern-0.5em /}

\def\gammax{ \gamma ^{\hat x } }
\def\gammat{ \gamma ^{\hat t } }

\def\psidagger { \psi ^{\dagger} }

\begin{document}

%%%%%%%%%%%%%%%% title page %%%%%%%%%%%%%%%%%%%%%%%%%%%%%%%%%%%%

\begin{titlepage}
\vfill
\begin{flushright}
hep-th/0308173\\
\end{flushright}

\vskip 1.5in

\begin{center}
\baselineskip=16pt
{\Large\bf  A Positivity Theorem for Gravitational Tension}
\vskip 1.0cm
{\Large\bf in Brane Spacetimes}
\vskip 0.5in
% {\large {\sl }}
% \vskip 10.mm
{\bf  Jennie Traschen} \\
%\\[2mm]
\vskip 1cm
%\vfill
{

       Department of Physics\\
       University of Massachusetts\\
       Amherst, MA 01003\\
       \vspace{6pt}
\texttt{traschen@physics.umass.edu}

}
\vspace{6pt}
\end{center}
\vskip 0.5in
\par
%\begin{center}
% {\bf ABSTRACT}
% \vskip 0.3in

\begin{abstract}

We study transverse asymptotically flat spacetimes without horizons that  arise from
brane matter sources. We assume that asymptotically there is a spatial translation Killing
vector that is tangent to the brane. Such spacetimes are characterized by a tension,
analogous to the ADM mass, which is a gravitational charge associated
 with the asymptotic spatial translation Killing vector.  Using spinor techniques,
we prove  that the purely gravitational contribution to the spacetime tension is positive definite.

\end{abstract}

%\end{center}
% \vskip 0.3in
% \begin{quote}
% 
% \vfill
% % \hrule width 5.cm
% \vskip 2.mm
% \end{quote}
\end{titlepage}

%%%%%%%%%%%%%%%%%%%%%%%%%%%%%%%%%%%%%%%%
\section{Introduction}

In this paper, we will study positivity properties of the gravitational mass and tension 
of $p$-brane spacetimes, using the 
spinorial methods of reference \cite{Witten:1981mf}.  
A $p$-brane spacetime is not asymptotically flat, but rather `transverse asymptotically flat',
 {\it i.e.}  as one approaches infinity in 
spatial directions transverse to the brane, the $n$-dimensional spacetime becomes $(n-p)$-dimensional
flat spacetime cross a $p$-dimensional space which is `tangent to the brane'. 
The tension of a p-brane spacetime \cite{Traschen}\cite{Townsend} is defined in close analogy to the mass of an 
asymptotically flat spacetime.  The gravitational contribution to the tension can be motivated as follows.  
A particle-like object, such as a billiard ball, has a rest mass.
If this mass becomes large, then self-gravity becomes important, and there are 
gravitational contributions to the total mass.  Now, consider an  elastic band.
The elastic band has a tension as well as its mass per unit length. Suppose the band is
either infinitely long, or wraps around a noncontractable loop.
 We can ask if there
is a gravitational contribution to this tension as well?  
The ADM mass of an asymptotically
flat spacetime is defined by a boundary integral of the long range gravitational field, and can
be constructed as a conserved charge associated with the asymptotic time translation Killing
vector $\partial/\partial t$.  In analogy with mass, one would expect
the spacetime tension to be a conserved charge associated with an asymptotic spatial translation
Killing vector $\partial/\partial x$ which is parallel to the brane.
This notion of spacetime tension for $p$-brane spacetimes was formalized in references
\cite{Traschen}\cite{Townsend}.  

Reference
\cite{Traschen} extended the Hamiltonian construction 
of gravitational charges \cite{Regge:1974zd} 
in a natural way to $p$-brane spacetimes.  The procedure is roughly as follows.
Let ${\cal H}_t$ be the
Hamiltonian that generates the flow along a timelike vector field asymptotic to
$\partial/\partial t$.  Regge and Teitelboim \cite{Regge:1974zd}\cite{Teitelboim} calculated the
variation of the Hamiltonian $\delta {\cal H}_t$ under variations of the dynamical degrees of 
freedom. The variations must
preserve the asymptotically flat boundary conditions.  In order that $\delta {\cal H}_t$ vanish on solutions to the 
equations of motion, they found that a certain surface term must be added
to the usual volume form of the Hamiltonian.
This surface term is a linear combination of the ADM gravitational charges.  The coefficients
of the charges are the asymptotic values of 
the Lagrange multipliers appearing in the definition of the Hamiltonian, and asymptote to global Poincare transformations.

Now consider a $p$-brane spacetime.  We assume that at
 transverse spatial infinity, there is a spatial Killing field 
$\partial/\partial x$, tangent to the brane.
 The covariance of the theory ensures that one can also construct a Hamiltonian
${\cal H}_x$ that generates flow along a spacelike vector field asymptotic to
$\partial/\partial x$.
Calculation of the variation $\delta{\cal H}_x$, with a Lagrange multiplier asymptotic
to a translation in the $x$-direction, then gives a surface term proportional to the 
spacetime tension $\mu$.  
The resulting definition  of $\mu$ is given in equation (\ref{foxtension}) below\footnote{Reference \cite{Traschen} 
established a first law of black $p$-brane mechanics that relates variations in the tension $\delta\mu$ to variations in 
the other standard thermodynamic quantities.  The definition of the tension $\mu$
 is then obtained by integrating up the
expression for $\delta\mu$ given in \cite{Traschen}, as one does to obtain the ADM gravitational charges in the Hamiltonian 
approach of reference \cite{Regge:1974zd}. In reference \cite{Townsend} a covariant
Lagrangian approach \cite{Abbott:1982ff} was used to define the tension $\mu$ of a $p$-brane spacetime.  It is straightforward
to check that these two methods give equivalent results.}.  

 In the perturbative limit
the tension is given in terms of the brane stress-energy, by 
\be
\mu =- \int  T_{xx}\ dv,
\ee
$i.e.$ the tension is minus the pressure of the matter stress-energy. This
 coincides with our intuitive definition of tension above.  
It is worth noting that, the tension $\mu$ is not the same as the ADM momentum 
in the $x$-direction, which is given in the perturbative limit by $P^x=\int T_{xt}\ dv$. 
 
In this work  we will show that the tension can be written as a contribution from the
matter stress-energy $T_{ab}$ and a purely gravitational contribution. As in the case
of gravitational mass, it is natural to ask if the tension is sign definite.
Using techniques similiar to those of Witten and Nester in proving the positive mass theorem
\cite{Witten:1981mf} \cite{nester},
we prove  that the purely gravitational contribution to $\mu$ is positive definite. 
The first step is to define the spacetime tension $\mu$ in terms of the long range gravitational field and
a spinor field, and show that the spinor definition of $\mu$
 agrees with the previously given definition. If the spacetime is static throughout
 the expression for $\mu$ simplifies. Further, we see that if a
 static spacetime has a local boost invariance in the $x-t$ directions, then
then the mass per unit length
minus the tension just depends on stress-energy terms. Horizons add a significant technical
complication. Since the proof uses Stokes Theorem, a horizon introduces an additonal 
boundary term, which one would like to show is geometrical, or nonnegative. In the case of
the mass, reference \cite{Gibbons:jg} extended Witten's proof to include horizons, by proving that
the horizon boundary term is zero. We will leave the issue of horizons and tension to
future work.

\section{Proof of positivity theorems}

Let ${\cal M}$ be a spacetime with $n$ total spacetime dimensions. ${\cal M}$ 
is assumed to have the topology of an $(n-p)$-dimensional noncompact
spacetime which is asymptotically flat, cross a $p$-dimensional space which may be compact or
noncompact.   In the asymptotically flat region, we assume that 
there is (at least) one spatial-translation Killing vector of the $p$-dimensional space.
Further,
we assume that in the asymptotic region, the spacetime has a covariantly constant spinor $\psi _0$.
The spacetime tension $\mu$ will be defined in terms of the long range gravitational field and
a spinor field which approaches $\psi _0$.

Assume that ${\cal M}$ can be foliated by spacelike slices $V_t$ with unit timelike normal $n_a$ 
and additionally that it can also be foliated by timelike slices $V_x$
with unit spacelike normal $x_a$.  
These surfaces can be used to define coordinates $(t,x)$ by defining
the  $V_t$ and $V_x$ to be surfaces of constant $t$ and $x$ respectively.
At transverse spatial infinity, we assume that $n^a$ and $x^a$ are proportional to the asymptotic time and space 
translation Killing vectors respectively.
We will further assume that the spacelike surfaces $V_{tx}$ with both $t$ and $x$ held constant are submanifolds.
These foliations allow us to write the metric on ${\cal M}$ in the following three forms
\bea\label{metric}
g_{ab}&=&x_a x_b +s_{ab}\\
& = & -n_a n_b+h_{ab} \\&=&  -n_a n_b +x_a x_b +q_{ab}.
\eea
Here $h_{ab}$ is the Euclidean metric on the hypersurfaces $V_t$,
$s_{ab}$ is the Lorentzian metric on the hypersurfaces $V_x$, and $q_{ab}$ is the Euclidean
metric on the submanifolds $V_{tx}$.  
The matrices $h^a_{\ b}$, $s^a_{\ b}$ and $q^a_{\ b}$ then act on vectors 
as projection operators onto the tangent spaces of the 
surfaces $V_t$, $V_x$ and $V_{tx}$ respectively. 
In the calculations below, we make use of a frame field $e^{\hat a}_{\ b}$, chosen such that
$e^{\hat t\, a}=n^a$ and $e^{\hat x\, a} =x^a$.
Our construction will take place in a volume $\Omega \subset {\cal M}$ that is bounded by initial and final spacelike 
slices $V_i$ and $V_f$, and by an outer boundary $\partial\Omega_\infty $ at infinity in the
transverse asymptotically flat directions.  We will assume that there are no
horizons in ${\cal M}$ and correspondingly that the volume $\Omega$ has no inner boundary. 
  
\subsection{A Modification of the Positive Energy Theorem }
We first present a modification of
 Witten's proof of the positive energy theorem \cite{Witten:1981mf}
as reformulated by Nester \cite{nester}. Our construction gives the purely gravitational
contribution to the mass in terms of the metric $q_{ab}$,
rather than $h_{ab}$ as in the original proofs. Both $q_{ab}$ and $h_{ab}$ are positive definite, so either
construction gives a positivity result, when the spinor field satisfies the Witten-Dirac 
equation defined in terms of  $q_{ab}$ or $h_{ab}$ respectively. The distinction is that
 $q_{ab}$ is the metric on a co-dimension two slice $V_{xt}$ and $h_{ab}$ is for a co-dimension
 one slice $V_t$. In the case of the spacetime tension $\mu$, it is necessary to reduce
  by two dimensions to
 prove positivity. We illustrate the technique first for the mass $M$. This also gives
  an expression for the mass
that is more useful for comparision to the tension.

 The Nester $2$-form is defined by
\be\label{bab}
B^{ab} = \psi^{\dagger} \gamma^{\hat t} \gamma^{abc} \nabla_c\psi ,
\ee
where $\nabla_a$ is the covariant derivative operator compatible with the full spacetime metric $g_{ab}$ and $\psi$ 
is a Dirac spinor.
If we define the vector $\xi ^a =-\psidagger \gammat \gamma ^a \psi$ and require that the 
spinor field approach a constant spinor $\psi_0$ at infinity,
then the ADM energy-momentum vector $P^a$ is given in terms of an integral over the
$(n-2)$-dimensional boundary  $(\partial V_t )^\infty$ at transverse spacelike 
infinity in a $p$-brane spacetime according to
\bea\label{admmass}
- P^a\xi_a &= &{1\over 16\pi}\int_{(\partial V_t )^\infty}
dS_a\left(B^{ab}+B^{ab\, \star}\right)n_b \\
& = &  -{1\over 8\pi}\int_{(\partial V_t )^\infty}
 dS_a ( \psi ^\dagger h^a_{\ d} h^b_{\ e} \gamma ^{de}\nabla _b \psi + c.c. ) 
\eea
In particular, if we take the asymptotic value of the spinor $\psi_0$ to be an eigenfunction of $\gammat$, then at infinity
$\xi^a =(\partial /\partial t)^a$ and $P^a\xi _a=-M$.  In order to focus on the relationship between the mass and the 
tension, we will assume such a choice for $\psi_0$ below. 

The construction proceeds by using Stokes theorem to rewrite the boundary integral
 (\ref{admmass}) as a volume integral.
Making use of the identity
$\gamma ^{abc} \nabla _a \nabla _c
\psi = -{1\over 2} G^{bc} \gamma_c\psi$, where $G_{ab}$ is the Einstein tensor and 
the decomposition of the spatial metric $h_{ab}= q_{ab} +x_a x_b$, the volume integrand
can be written as
\bea\label{massone}
n_b \nabla_a B^{a b}
&= & -{1\over 2} n^aG_{a c}  \psi \gamma ^{\hat t}\gamma^ c\psi
 +\nabla_a \psi^{\dagger} q^{ab}  \nabla_b \psi
  -(\nabla_a \psi^{\dagger}q^a _c\gamma ^c ) ( q^b _d\gamma ^d  \nabla_b \psi ) \\
& & -x^a \nabla_a \psi^{\dagger} \gammax  (\gamma^b q^{\ c}_b \nabla_c \psi )
-(\nabla _a \psi ^{\dagger} q^a_{\ b} \gamma ^b ) \gammax x^c\nabla _c\psi
 \nonumber
\eea
Now suppose that in the region $\Omega$, the spinor $\psi$ satisfies a modified Dirac-Witten equation given by
\be\label{dirac}
q^c_{\ b} \gamma^b \nabla _c \psi =0 .
\ee
Then the third, fourth, and fifth terms on the right hand side of (\ref{massone}) are zero.
Equation (\ref{dirac}) is a differential equation defined on each of the slices $V_{tx}$. So,
the solution $\psi$ will depend on the coordinates $x$ and $t$ as parameters. 
We assume that this dependence is sufficiently smooth to allow us to use Stokes theorem below.
If we now integrate equation (\ref{massone}) over any of the hypersurfaces $V_t$, 
then using Einstein's equations $G_{ab} = 8\pi T_{ab} $, the result is
\be\label{masstwo}
M ={1\over 8\pi }\int _{V_t} \sqrt{h}\ (8\pi T_{ab} n^a \xi ^b 
+2\nabla _a \psidagger q^{ab} \nabla _b \psi ).
\ee
This gives an alternative demonstration that the ADM mass is positive if the stress-energy satisfies the
dominant energy condition. 

In the original Witten-Nester construction, the metric $h_{ab}$, rather than $q_{ab}$,
 appears in the first line of equation (\ref{massone}), and the second line of (\ref{massone}) does
 not arise at all.  The spinor field must be
 a solution to $h^c_{\ b} \gamma^b \nabla _c \psi =0$. Let us compare the Witten-Nester
 construction to the present case. We are
  able to reduce by two dimensions and still prove positivity,
 because the extra terms generated are still proportional to the Dirac equation (\ref{dirac}).
 It is reasonable to ask if one could reduce further, $i.e.$, let $g_{ab}= L_{ab} +q_{ab}$
 where $ L_{ab} ,q_{ab}$ are orthogonal, and $q_{ab}$ is positive definite. Can the expression
 for $M$ be written just in terms of the transverse metric $q_{ab}$? Starting with the
 two-form $B^{ab}$ the answer is no, at least with the requirment that $\psi$ satisfy
 a Dirac equation. Cross terms arise which are not zero or sign definite. It would be
 interesting to know if starting with a higher dimension form would yield a positiviy
 proof for the mass per unit length of a general $p$-brane.

\subsection{Positivity theorem for the gravitational component of tension}
We want to ask whether, or not, there is a similar positivity result associated with the tension of a $p$-brane spacetime.
We start by defining a Nester form $E^{ab}$ in analogy with equation (\ref{bab}), but with $x^a$ assuming the special role
played by the time direction in the positive energy construction above.  Let
\be\label{eab}
E^{ab} = \psi^{\dagger} \gamma^{\hat x} \gamma^{abc} \nabla_c \psi .
\ee
Then it is straightforward to show that
\bea\label{spinoridentity}
x_b \nabla_a  E^{a b} & = & -{1\over 2} x^a G_{a  c} \psidagger\gamma ^{\hat x}\gamma^ c\psi
 +\nabla_a \psi^{\dagger} q^{ab}  \nabla_b \psi
  -(\nabla_a \psi^{\dagger}q^a _c\gamma ^c ) ( q^b _d\gamma ^d  \nabla_b \psi ) \\
  & &
  -\nabla_a \psi^{\dagger}n^a \gammat ( \gamma^b q^{\ c} _b \nabla_c \psi )
-(\nabla _a \psi ^{\dagger} q^a_{\ b} \gamma ^b )\gammat n^c\nabla _c\psi .
\nonumber
 \eea
If the spinor $\psi $ satisfies the Dirac-Witten equation (\ref{dirac}), then again
only the first two terms on the
right hand side of equation (\ref{spinoridentity}) are nonzero.
Stokes theorem can then be used to integrate (\ref{spinoridentity}) over a Lorentzian slice $V_x$.
 The boundary of this slice 
$\partial V_x$ includes initial and final slices $V_{x}^{i}$ and $V_{x}^{f}$, as well as
the boundary at transverse spatial infinity. 
  However, direct computation shows that
if $\psi$ solves the Dirac equation, then the quantity $n_a x_b E^{ab} =0$.  Therefore
  $V_{x}^{i}$ and $V_{x}^{f}$
do not contribute to the tension.   The boundary at infinity is a product of a spatial surface that we denote 
$(\partial V_{tx})^{\infty}$ and a finite time
interval $\Delta t$ between initial and final time slices.  Because the spacetime is transverse asymptotically flat, 
the integrand at infinity is independent of the time coordinate $t$ and the integral is proportional to $\Delta t$. 
This leads to an expression for the tension $\mu$ per unit time 
as the boundary integral
\bea\label{tensiondef}
\mu &=& {1\over  8\pi(\Delta t ) }\int _{(\partial V_x ) ^\infty} 
dt da_b \left( E^{bc}+ E^{bc\ *}\right)x_c\\
&=&- \int _{(\partial V_{tx})^{\infty}}  da_a s^a_{\ d} s^b_{\ e} (  \psi ^\dagger 
\gamma ^{de}\nabla _b \psi +c.c. ).
\eea
It turns out that the definition of tension given in equation (\ref{tensiondef}) is the
same as previously given, without reference to spinor fields,
 in \cite{Traschen} and \cite{Townsend}. The argument is a bit detailed, and we defer it
 until the end of the section.
 
 Using Stokes theorem, then on solutions to the Dirac equation equation
 (\ref{spinoridentity}) becomes
\be\label{positive}
8\pi\mu  =
{1\over  8\pi(\Delta t ) } \int _{V_x} \sqrt{-s}
  \left ( -8\pi T_{ab} \chi^a x^b + 2 q^{ab}\nabla_a\psi^\dagger \nabla_b\psi \right )
\ee
where $\chi ^c ={1\over 2}(\psidagger\gamma ^{\hat x}\gamma^ c\psi +
 \psidagger\gamma^ {c \dagger} \gammax\psi)$.
  When $\psi _0$ is chosen to be an eigenfunction
of $\gammat$, then the vector  $\chi ^c $ approaches$ {\partial\over \partial x}$
 at infinity. Equation (\ref{positive}) shows that
  the purely gravitational contribution to the tension, the second term in the
 integrand, is positive definite.

If the spacetime is static with timelike Killing vector $l^a ={\partial \over \partial t}$ everywhere,
 the volume integral in equation (\ref{positive}) can be reduced to an integral over a spacelike slice $V_{tx}$. 
 Choose a slice $V_{tx}^i$, and foliate $\Omega$
with surfaces $V_{tx}$ by moving $V_{tx}^i$ along the integral curves of $l$. The volume integrand
is independent of the symmetry coordinate $t$, and (\ref{positive}) becomes
\be\label{positivetwo}
8\pi\mu  =
 \int _{ V_{tx}} \sqrt{-s}
  \left ( -8\pi T_{ab} \chi^a x^b + 2 q^{ab}\nabla_a\psi^\dagger \nabla_b\psi \right )
\ee
This relation illustrates why
we have called $\mu$ the tension. The stress-energy contributes to $\mu$ as one would 
expect for a tension. For example,
suppose the stress energy is a perfect fluid. Since
the spacetime is static, the velocity of the fluid $u^a$ , is proportional to $l^a$. Choose
the slices so that $n^a =u^a$. Then $T_{ab} \chi^a x^b =
\psidagger \psi T_{\hat x \hat x}$, so that
 the stress energy term in (\ref{positivetwo}) is minus the pressure in the $\hat x -$direction, times
the positive quantity $\psidagger\psi$, which approaches one at infinity. For comparision,
in this same example, the stress energy contribution to $M$ is $\psidagger \psi T_{\hat t \hat t}$.
Equations (\ref{positive}), (\ref{positivetwo}) constitute our main results.

We can add some intuition to the definition of $\mu$ by recalling a previous result  \cite{Traschen}.
Let $g^0 _{ab}$ have the spatial translation Killing field $({\partial \over \partial x})^b$
 everywhere, and let $g_{ab}$ be perturbatively close to $g^o _{ab}$. We showed that
for arbitrary perturbations, $ \delta\mu =-\int  \delta T_{ab} x^a
({\partial \over \partial x})^b$, which is how one expects the stress-energy to contribute to
the tension. For comparision, the perturbation to the mass satisfies
 $ \delta M = \int  \delta T_{ab} n^a({\partial \over \partial t})^b$. 
 These results agree with the perturbative limits of the spinor relations
 (\ref{positivetwo}) and (\ref{masstwo}) respectively, though some care must 
 be taken in verifying this, since the spinor field receives first order corrections
 and this changes the associated vector.
 If the background spacetime $g^0 _{ab}$ contains a black brane horizon, then both $\delta M$
 and $\delta\mu$ receive contributions from the boundary at the horizon. For the mass, this
 is the First Law,  $ \delta M = {1\over 8\pi} \kappa\delta A +
 \int  \delta T_{ab} n^a({\partial \over \partial t})^b$. For the tension,
 we derived the relation \cite{Traschen}
\be\label{black}
 \delta\mu = {1\over (n-2) 8\pi }\kappa \delta A 
-\int  \delta T_{ab} x^a ({\partial \over \partial x})^b
\ee
where $A$ is the area of the horizon and $\kappa$ the surface gravity.
 A topic for future work is to investigate the extension of the spinor
 result (\ref{positivetwo}) to spacetimes with horizons.

Lastly, we return to the demonstration that equation (\ref{tensiondef}) does indeed
agree with the definition of the tension $\mu$ given in references
\cite{Traschen} and \cite{Townsend}. 
In the asymptotic region we can write the metric as 
$g_{ab}=g^0_{ab}+ \delta g_{ab}$, where the background metric
$g^0_{ab}$ is flat with periodic identifications in the directions along the brane\footnote{If the compact
space is not flat, for example, a Calabi-Yau space cross an $S^1$, there are additional terms in
equation (\ref{tensiontwo}) below.}. Similiarly, $s_{ab}=s^0_{ab}+ \delta s_{ab}$,
$q_{ab}=q^0_{ab}+ \delta q_{ab}$. Let $V^a$ be the spatial translation Killing field of $g^0_{ab}$.
 The expression for $\mu$ given in reference \cite{Traschen} is 
\be\label{foxtension}
\mu =
   {1\over 16\pi}\int da _b [ \tilde{D} _a (\delta s_{cd}g^{ac} g^{bd} ) -
\tilde{ D} ^b (\delta s_{cd}g^{cd} ) ] V ^{\hat x},
\ee
where $\tilde D$ satisfies $\tilde {D}_a s^0_{bc}=0$. 

The definition (\ref{tensiondef}) must be rewritten in terms of the perturbation to the metric.
  The boundary integral in equation (\ref{tensiondef}) depends on the derivative operator
$\nabla_a$.  In the asymptotic regime it is useful to write the derivative operator as 
$\nabla_a=\nabla^0_a + (\nabla_a- \nabla^0_a)$, where
$\nabla_a^0$ denotes the flat derivative operator of the flat background metric $g^0_{ab}$.   
If we now let $E^{ab}[\nabla^0]$ be the Nester form defined
with respect to the flat derivative operator, then the quantity $E^{ab}[\nabla^0]\, x_a$ can be shown to vanish at infinity.  
In order to evaluate the boundary term in (\ref{tensiondef}), it then suffices to compute the quantity
$b^a = q^a _{\ b}\, E^{bc} [\nabla -\nabla ^0]\, x_c $ in terms of the perturbation $\delta g_{ab}$ in the 
asymptotic regime. The operator $(\nabla_a- \nabla^0_a)$ acting on a tensor field is given in terms of the connection coefficients 
\be
\Gamma ^d _{ce} =
{1\over 2} g^{db}\left(\nabla ^{0} _c\, \delta g_{be}+
\nabla ^{0} _e\, \delta g_{bc} - \nabla ^{0} _b\, \delta g_{ce} \right).
\ee
Making use of this result, we find that
\be\label{baandba}
 b^a + b^{a *}  =
{1\over 4}\left( q^a _{\ b}  \psidagger\{ q^d _{\ c} (\gamma ^{bc}\gamma ^{mn}
-\gamma ^{mn\dagger} \gamma ^{bc} )
+n^d \gammat (-\gamma^b \gamma ^{mn}  +  \gamma ^{mn}\gamma ^b ) \}\psi\right)\  \nabla _{[m} \delta g_{n] d}.
\ee
In order to proceed further we
must specify the rates at which the different components of
 $\delta g_{ab}$ vanish at transverse spatial infinity. 
Letting $i,j\neq x,t$,  we assume that the components $\delta g_{tt}$,
 $\delta g_{xx}$ and $\delta g_{ij}$ fall off as  $1/r^{n-p-3}$ and
that the components $\delta g_{ti}$, $\delta g_{tx}$ and
  $\delta g_{xi}$ fall off at a faster rate.  These latter conditions 
imply, in particular, that the extrinsic curvatures of
 the slices $V_{t}$ and $V_x$ fall off faster than $1/r^{n-p-3}$.
Calculations then lead to useful intermediate results.  In particular, it follows that 
\be
q^{am}n^e  \gamma ^n \nabla _m \delta g_{ne}\simeq
 q^{am}\nabla _m \gamma^{\hat t}\delta g_{\hat t \hat t}
% \gammat + h.o.t.'s
\ee
where the equality holds up to terms which vanish in the limit that the
boundary integral goes to infinity.  Similarly, it follows that
\be 
q^{\ e} _c  \gamma ^{mn} \nabla _{[m} \delta g_{n] e} \simeq
  q^{\ e} _c   \gamma ^{mn} D _{[m}  \delta q_{n] e} ,
\ee
where $D_a$ is the derivative operator on the surfaces $V_{tx}$ satisfying $D_a q_{bc}=0$.  
Putting these results together allows one to rewrite the tension boundary term in equation (\ref{tensiondef}) as
\be\label{tensiontwo}
\mu  = {1\over 16\pi}\int da _b  [ D _a (\delta q_{cd}g^{ac} g^{bd} )
- D ^b (\delta q_{cd}g^{cd} ) + D^b \delta g_{\hat t \hat t} ] \psidagger _0\psi _0 
 \ee
Use $q^a _b D_a \delta q_{cd} \simeq q^a _b \tilde{ D} _a (\delta s_{mn}q^m _a q^n _b)$ and
$\delta s_{cd} g^{cd} = \delta q_{cd} g^{cd} -\delta g_{\hat t \hat t}$. Then the boundary
term in equation (\ref{foxtension}), given
 in terms of $\delta s _{ab}$, agrees with  equation (\ref{tensiontwo})
 with the identification $V ^{\hat x} = \psi _0 ^\dagger \psi _0 $.

\subsection{ Comparing Mass and Tension}

One notices an interesting feature when comparing the spinor expressions for the mass and tension
in static spacetimes.
Equations (\ref{masstwo}) and (\ref{positivetwo}) show that the purely gravitational
 contribution to the mass and the tension are the same when the volume elements
 are equal. In such a case the difference between the mass and the tension is only due
  to the stress energy terms. One situation where this occurs, noted in \cite{Townsend}, is when
  ${\partial\over \partial x}$ is a Killing vector throughout the spacetime, $and$
  the spacetime is (globally) boost-invariant in the $\hat x  -\hat t  $ plane. 
  Then the symmetry implies that the boundary terms for $M$ and $\mu$ are the same.
  To see that there are less restrictive cases,
   choose the timelike  normal to be parallel to the static Killing field,
$n_a =-N\nabla _a t $. The volume elements are equal if $x_a$ can be chosen such that
 $x_a =N\nabla _a x$, for example, if the metric can be put in the form
$ ds^2 = N^2 (x, x^j ) ( -dt^2  +dx^2 ) +q_{ij}dx^i dx^j $. That is,  ${\partial\over \partial x}$
need not be a Killing vector throughout the spacetime, but there is a local boost invariance
in the $\hat x  -\hat t  $ plane.
 Physically, one would expect this to result from a source with 
   $p_x = -\rho$ locally, but that the pressure depends on position along the string.
   
A related issue is whether, or not, there are solutions which are static but not translationally
invariant. An example of such a solution has been conjectured in analysis of
the Gregory-Laflamme instability of the black string \cite{Gregory:vy}. It is 
conjectured that (a) the unstable horizon evolves to a static, translationally non-invariant
   black string \cite{Horowitz:2001cz} \cite{Horowitz:2002ym},
   and that (b) at the onset of instability
   there is also a static translationally non-invariant state \cite{Gubser:2001ac} \cite{Reall:2001ag}.
Our present results 
do not apply to the black string
since we have not included horizons. So, as far as the above spinor expressions apply,
we are interested in  smooth fill-ins of the conjectured solutions
(a), (b). This leads to an
interesting question. An $x$-dependent pressure means that there is a pressure gradient
along the brane, which usually makes things flow, and would contradict the assumption
of staticity. On the other hand, a star has a radial pressure gradient. The star can
still be static because gravity is attractive. So it may be that the positive tension from
the gravitational field can balance a pressure gradient. Then one can understand the 
positivity of the gravitational tension as a reflection of the attractiveness of gravity.
Perhaps the non-symmetric static black-string 
solutions are limiting cases of this sort of equilibrium.

 \bigskip
\noindent
{\bf Acknowledgements:} I would like to thank David Kastor and Subir Mukhopadhyay for useful
conversations, and the Kavli Institute for Theoretical Physics and the Aspen Center for Physics for their
hospitality.  This work was supported in part by NSF grant PHY-0244801.

\end{document}